\documentclass[twocolumn,amsmath,amssymb,aps,pra]{revtex4}
\usepackage{bm}
\usepackage{amsfonts}
\usepackage{amsmath}
\usepackage{graphicx}
\usepackage{float}

\begin{document} 
\title{Knotted nodal lines in superpositions of Bessel-Gaussian light beams}
\author{Tomasz Rado\.zycki}
\email{t.radozycki@uksw.edu.pl}
\affiliation{Faculty of Mathematics and Natural Sciences, College of Sciences, Institute of Physical Sciences, Cardinal Stefan Wyszy\'nski University, W\'oycickiego 1/3, 01-938 Warsaw, Poland} 
\begin{abstract}
A simple analytical way of creating superpositions of Bessel-Gaussian light beams with knotted nodal lines is proposed. It is based on the equivalence between the paraxial wave equation and the two-dimensional Schr\"odinger equation for a free particle. The $2D$ Schr\"odinger propagator is expressed in terms of Bessel functions, which allows to obtain directly superpositions of beams with a desired topology of nodal lines. Four types of knots are constructed in the explicit way: the unknot, the Hopf link, the Borromean rings and the trefoil. It is also shown, using the example of the figure-eight knot, that more complex structures require larger number of constituent beams as well as high precision both from the numerical and the experimental side. A tiny change of beam's intensity can lead to the knot ``switching''.
\end{abstract}
\maketitle

\section{Introduction}\label{int}

In recent decades, it has become possible to analytically ``design'' and experimentally generate beams of light with given topological properties. This marriage of optical phenomena and topology has led to the birth of the so-called topological optics. Topological effects have entered into light propagation in several instances.
One intensively studied family constitute phase singularities of helical character. Such waves, apart from spin, are also endowed with orbital angular momentum. Particular interest of researchers was attracted here by  Laguerre-Gaussian~\cite{lg,lg2,arlt2} or Bessel~\cite{arlt2,durnin1,durnin2,vg,ibb1,tr3} beams possessing a property of vorticity and an associated ``charge'' or topological index. 

An example that can also be classified as topological are polarization singularities connected with the inability to fully specify the polarization at certain places~\cite{nye,nyh,freu,ddd,car}. Yet another idea is that of the ``knotted'' light. This primarily involves electric or magnetic field lines which can get entangled~\cite{ran,ir,besi,db,kedia,arr,ho,arr2,kpi}, but also the nodal lines of wave intensity or, in other words, optical vortex lines which can develop topologically non-trivial structures both in exact and paraxial regimes~\cite{bd,den,bd2,bkj,kle,deklerk,su}. This latter case covering some kind of {\em doubly-topological} structures (i.e., knotted vortices, which per se are topological entities mentioned above) is our main concern in this work. From an experimental point of view, these types of beams can be produced, which opens up many research and application possibilities~\cite{leach,leach1,sha,wil}.

Quite paradoxically, while dealing with light beams it is not bright regions but those of darkness, where the exact destructive interference occurs, that are of main interest. They are connected with the presence of phase singularities, as stated above, since the phase is undetermined for vanishing field. Such areas of suppressed intensity can serve as traps for both polarizable neutral particles with negative polarization constant (such as blue-detuned atoms)~\cite{dav,odtna,sheng,frie,trb}, for charged particles like electrons through the ponderomotive potential~\cite{ibb1}, and even for micrometer-sized objects. In case of neutral atoms the gradient forces arising from the inhomogeneities of the electric field occur due to the Stark effect~\cite{dk} and for larger objects the trapping appears via Mie scattering~\cite{mie,neu,dho1,die}.

Generating nodal lines with highly non-trivial geometrical properties and studying knotted structures as such, can prove useful for instance for manipulating particles using this mechanism of trapping and guiding or through implementations into other physical systems. Many potential practical applications range from physics through chemistry to biology or medicine~\cite{liu,ste,fazal,pad,woe,bowpa,grier1,tka,mar,hall,brad,dan}. 

The question of constructing a kind of a knotted trap for silica spheres was undertaken in~\cite{sha} and for neutral atoms in~\cite{trc}. In this latter work we proposed a simple method to design ``arbitrary'' traps of this sort from the superposition of simple Gaussian beams. It was also shown in the numerical way that the trajectories of neutral polarizable particles are really confined on these knotted nodal lines. 
However, in principle this kind of traps can be made of more complex beams, as for instance Laguerre-Gaussian (LG) beams or Bessel-Gaussian (BG) beams for which some knotted nodal lines are successfully obtained~\cite{bd1,bd3,lea,king,pad2}. Therefore, in the present work we would like to extend to BG beams the method advocated in our previous paper~\cite{trc}. From the experimentalist point of view it seems essential that knots could be formed of a variety of beam types. Contrary to some earlier attempts the present approach should allow, in principle, to construct in an easy and straighforward way any knots (composed of nodal lines) that can be obtained from the so-called Milnor polynomial. Moreover, the BG beams are already routinely achievable in experiments. The procedure leading to this result will be discussed in detail in Section~\ref{genpro}.  

When dealing with laser beams it has proved convenient to introduce dimensionless coordinates through the relations
\begin{equation}\label{diml}
\xi_x=k x,\;\;\;\xi_y=k y,\;\;\;\xi=\sqrt{\xi_x^2+\xi_y^2} ,\;\;\;\zeta=k z.
\end{equation}
This corresponds to measuring distances in the units $k^{-1}=\lambda/2\pi$ and ensures that our formulas become as simple as possible. It should be noted that for the third component the symbol $\zeta$ instead of $\xi_z$ is used. This is due to the fact that this coordinate plays some special role in our considerations, i.e. that of the ``time'' in the corresponding Schr\"odinger equation. Therefore, in what follows the bold mathematical symbols stand for only two-dimensional vectors (i.e., $\bm{\xi}=[\xi_x,\xi_y]$ and $\bm{r}=[x,y]$).

A real wave close to the propagation axis satisfies the so called paraxial equation which in the dimensionless coordinates has the form:
\begin{equation}\label{paraxial}
\mathcal{4} \Psi(\bm{\xi},\zeta)+2i\partial_\zeta \Psi({\bm{\xi}},\zeta)=0,
\end{equation}
with $\mathcal{4}$ representing the two-dimensional Laplace operator. In order to construct superpositions of BG beams on one hand satisfying Eq.~(\ref{paraxial}) and on the other displaying a given topological structure the clear similarity of~(\ref{paraxial}) to the Schr\"odinger equation for a free particle in $2D$ will be exploited as in~\cite{trc}. The latter equation has the form:
\begin{equation}\label{schr}
-\frac{\hbar^2}{2m}\,\mathcal{4}\Psi(\bm{r},t)=i\hbar\partial_t\Psi(\bm{r},t),
\end{equation}
and becomes identical to~(\ref{paraxial}) upon the the identification:
\begin{equation}
mc^2=\hbar\omega,\;\;\;\; \zeta=c t.
\label{id}
\end{equation}
This allows the well-known Schr\"odinger propagator to be made use of, as described in general terms in the next section. In Section~\ref{skb} this general procedure is implemented in five subsequent examples: the unknot, the Hopf link, the Borromean rings, the trefoil and the figure-eight knot. In a systematic way, the concrete superpositions of coaxial BG beams are derived, which yield these nodal structures.

\section{General procedure}\label{genpro}

The two-dimensional free Schr\"odinger propagator may be written in terms of Bessel functions $J_n(x)$ with the use of our dimensionless variables~(\ref{diml}) as follows:
\begin{eqnarray}
&&K(\xi,\phi,\zeta;\xi',\phi',\zeta')=\frac{1}{2\pi}\,\sum\limits_{n=-\infty}^{\infty}
e^{in(\phi-\phi')} \nonumber\\
&&\;\;\;\;\times\int\limits_0^\infty \mathrm{d}t \, t J_n(t\xi)J_n(t\xi')e^{-i(\zeta-\zeta')t^2/2}.
\label{propagator}
\end{eqnarray}
One can verify by a direct calculations that the paraxial equation~(\ref{paraxial}) (or Schr\"odinger equation) is satisfied by the above expression. It is inessential whether $K(\xi,\phi,\zeta;\xi',\phi',\zeta')$ itself meets the condition of the paraxial approximation, which for certain function $f(\bm{\xi}, \zeta)$ in our units reads
\begin{equation}
\partial^2_\zeta f(\bm{\xi}, \zeta)\ll \partial_\zeta f(\bm{\xi}, \zeta),
\label{pax}
\end{equation}
since $K$ is only a tool for obtaining a beam envelope $\Psi(\bm{\xi}, \zeta)$ of the physical significance for our considerations. As we will see in the following sections, this envelope is virtually a superposition of BG paraxial modes.

When $\zeta\rightarrow \zeta'$, one obtains
\begin{equation}
\lim_{\zeta\rightarrow \zeta'}K(\xi,\phi,\zeta;\xi',\phi',\zeta')=\delta^{(2)}(\bm{\xi}-\bm{\xi}'),
\label{link}
\end{equation}
as it is expected. All one needs is to apply the identity
\begin{equation}
\delta\left(\phi-\phi'\right)=\frac{1}{2\pi}\sum_{n=-\infty}^{\infty}e^{in(\phi-\phi')}
\label{deltaph}
\end{equation}
together with~\cite{arfken}
\begin{equation}
\delta\left(x-x'\right)=x\int_{0}^{\infty}\mathrm{d}t\, tJ_{n}\left(xt\right)J_{n}\left(x't\right),
\label{deltab}
\end{equation}
and then exploiting the property of the Dirac delta function in two dimensions:
\begin{equation}
\frac{1}{\xi}\,\delta(\phi-\phi')\delta(\xi-\xi')=\delta^{(2)}(\bm{\xi}-\bm{\xi}').
\label{polard}
\end{equation}
leads to (\ref{link}).

Now we move on to the construction of a knot of a certain required topology. In this work we focus on four chosen knotted structures: the unknot (or simple ring), the Hopf link, the Borromean rings and the trefoil and at the end it is shown using the example of the figure-eight knot what kind of complications occur for more complex windings. The details of the construction are the subject of the {\em knot theory} to be found elsewhere~\cite{bra,bkj,king,bode} and stay beyond the scope of this work. The main steps are listed below. 

A knot is in general a closed curve in $\mathbb {R}^3$ which constitutes a homeomorphic image of $S^1$. Contrary to our everyday's meaning of a knot, it forms a closed loop.  Typically, this curve is a nodal line of a certain complex-valued function of the spatial variables $x,y,z$ (or $\xi_x,\xi_y,\zeta$ in our case). If it is composed of several disjoint loops, that are tangled up, it is called a link, as the Hopf link for instance. The other example of the same family are the Borromean rings. Both of these cases, among others, will be dealt with in the next section.

The construction under consideration proceeds as follows. As a first step,  one creates a polynomial $q(u,v)$ of two complex variables $u$ and $v$ which satisfy the condition for the three-dimensional sphere: $|u|^2+|v|^2=1$. The examples of such polynomials are considered in the following section.  All the nodal points (i.e, those  where where $q(u,v)=0$) represent an algebraic knot. For the first three cases dealt with in the following section, these polynomials have the form:
\begin{equation}
q(u,v)=\prod\limits_{k=0}^{n-1} (u-\varepsilon_n^{(k)}v),
\label{qgen}
\end{equation}
with $\varepsilon_n^{(k)}$, $k=0,1,2,\ldots,n-1$, denoting the subsequent $n$th roots of unity.
Such a knots, however, would be located on $S^3$, which is important for their classification, but inappropriate for our purposes. We are concerned about knots in ${\mathbb R}^3$, so the next step is to exploit the stereographic projection by means of the relations:
\begin{subequations}\label{stereo}
\begin{align}
u(\bm{\xi},\zeta)&=\frac{\bm{\xi}^2+\zeta^2-1+2 i \zeta}{\bm{\xi}^2+\zeta^2+1},\label{stereou}\\
v(\bm{\xi},\zeta)&=\frac{2(\xi_x+i\xi_y)}{\bm{\xi}^2+\zeta^2+1}.\label{stereov}
\end{align}
\end{subequations}
The obtained equation $q(u(\bm{\xi},\zeta),v(\bm{\xi},\zeta))=0$ defines a knot curve as an intersection of two surfaces in three-dimensional space: $\operatorname{Re}q(u(\bm{\xi},\zeta),v(\bm{\xi},\zeta))=0$ and $\operatorname{Im} q(u(\bm{\xi},\zeta),v(\bm{\xi},\zeta))=0$. Recalling that $q(u,v)$ was a polynomial, $q(u(\bm{\xi},\zeta),v(\bm{\xi},\zeta))$ can again be treated as a polynomial (the so called Milnor polynomial~\cite{milnor}), upon removing the common denominator appearing in~(\ref{stereo}). The appropriate Milnor polynomials will be below denoted with $q_M(\bm{\xi},\zeta)$. These polynomials will constitute the basis for obtaining superpositions of BG beams with identical topology of nodal lines. The ``initial'' envelope $\Psi(\bm{\xi}, 0)$ will be created as a sum of such BG modes, that for small $\xi$ exhibit the same bahaviour as $q_M(\bm{\xi},0)$, i.e. the following replacement will be made: 
\begin{equation}
q_M(\bm{\xi},0)\mapsto\Psi(\bm{\xi}, 0)=e^{-\kappa\xi^2}\sum_{l,m}\alpha_{lm}e^{2im\phi} J_m(\chi_{lm}\xi),
\label{suma}
\end{equation}
together with the appropriate choice of the coefficients $\alpha_{lm}$. The summations with respect to $l$ and $m$ run over a range dictated by the form of a given Milnor polynomial for $\zeta=0$, and $\kappa>0$. It is outlined in detail in the next section. 
The values of coefficients $\alpha_{lm}$ are chosen so as to exactly (but apart from the Gaussian factor $e^{-\kappa\xi^2}$) reproduce the polynomial $q_M(\bm{\xi},0)$ as $\xi\ll 1$. On the other hand, the values of $\chi_{lm}$ can be set arbitrarily, dependent on the specific beams used in an experiment, and are connected with half-aperture of the appropriate beam cone (see formula~(\ref{nota2})). Consequently $\alpha_{lm}$'s which can be related to the beams' intensities, are functions of $\chi_{lm}$'s. Later, when considering specific examples, we prefer to denote the coefficients $\alpha_{lm}$ with $\alpha_l$, $\beta_l$, $\gamma_l$ and so on, in order to avoid too many indices. Usually, at least for the simplest knots, $m$ runs from 1 to say 2, 3 or maximally 4, so the symbols $\alpha_l$, $\beta_l$, $\gamma_l$, $\ldots$ are sufficient. For the same reason the symbols $\chi_{lm}$'s are modified in an obvious way.

Finally the full beam is found according to the formula
\begin{equation}\label{evol}
\Psi(\bm{\xi},\zeta)=\int d^2\xi' K({\bm \xi}, \zeta; {\bm \xi}',0) \Psi(\bm{\xi}',0)
\end{equation}
and automatically satisfies the paraxial equation~(\ref{paraxial}).

\section{Specific knotted beams}\label{skb}

In this section five concrete examples of knotted beams are dealt with. The suggested technique allows in principle to create BG beams with any knotted topology derived from a Milnor polynomial $q_M$. Yet, more complex knots require superpositions of larger number of mods, the structure of the nodal lines becomes shallower, and the computer time necessary to visualize them grows significantly.  For this reason, our analysis will be restricted to relatively simple knots, although -- apart from the first -- far non-trivial ones.

\subsection{The unknot}\label{uk}

The simplest knot, called {\em the unknot} is a simple ring. The polynomial $q(u,v)$, that yields such a ring, can be obtained from~(\ref{qgen}) when setting $n=1$:
\begin{equation}
q(u,v)=u-v.
\label{qring}
\end{equation}
Upon substitution of $u$ and $v$ as given in~(\ref{stereo}), the Milnor polynomial at $\zeta=0$ is found in the form
\begin{equation}
q_M(\bm{\xi},0)=-1+\xi_x^2+\xi_y^2-2(\xi_x+i\xi_y),
\label{fring}
\end{equation}
In order to preserve the capability of resizing the knot and adjust its dimension to the conditions of the paraxial approximation, a parameter $\gamma$ will be introduced, which can be called a ``scaling parameter'', thereby modifying  the form of the Milnor polynomial at $\zeta=0$ to
\begin{equation}
q_M(\bm{\xi},0)=-1+\gamma^2(\xi_x^2+\xi_y^2)-2\gamma(\xi_x+i\xi_y),
\label{fringg}
\end{equation}
without, however, altering the emerging-knot topology. Larger values of $\gamma$ lead to smaller knots. In the figures presented underneath, the value of this parameter is set to $10$, which results in typical knot sizes to be small fractions of a wavelength. For optical frequencies, they can be then called ``nano-knots''. 

Let us now replace~(\ref{fringg}) with
\begin{equation}
q_M(\bm{\xi},0)\mapsto \alpha_1 J_0(\chi_1\xi)+\alpha_2J_0(\chi_2\xi)+\beta e^{i\phi} J_1(\chi\xi)
\label{inta}
\end{equation}
according to the formula~(\ref{suma}), where instead of Cartesian $\xi_x,\xi_y$, the polar coordinates $\xi,\phi$ have been introduced and with the aforementioned and obvious renaming of constants $\alpha_{lm}$ and $\chi_{lm}$.  These latter values will, in general, be small since they define the aperture half-angles of the beam cones. 

Using now the well-known power expansion of the Bessel functions:
\begin{equation}
J_n(z)=\left(\frac{z}{2}\right)^n\sum_{k=0}^\infty\frac{(-1)^k(z/2)^{2k}}{k!(n+k)!}
\label{besn}
\end{equation}
it is easy to demonstrate that with the appropriate choice of the coefficients, the first few terms of the Maclaurin  expansion of expression~(\ref{inta}) in $\xi$ accurately reproduce $q_M(\bm{\xi},0)$ as given by the formula~(\ref{fringg}). This proper choice is as follows:
\begin{equation}
\alpha_1=\frac{\chi_2^2-4\gamma^2}{\chi_1^2-\chi_2^2},\;\;\;\; \alpha_2=\frac{\chi_1^2-4\gamma^2}{\chi_2^2-\chi_1^2},\;\;\;\; \beta=-\frac{4\gamma}{\chi}.
\label{aab}
\end{equation}

\begin{figure}[h]
\begin{center}
\includegraphics[width=0.45\textwidth,angle=0]{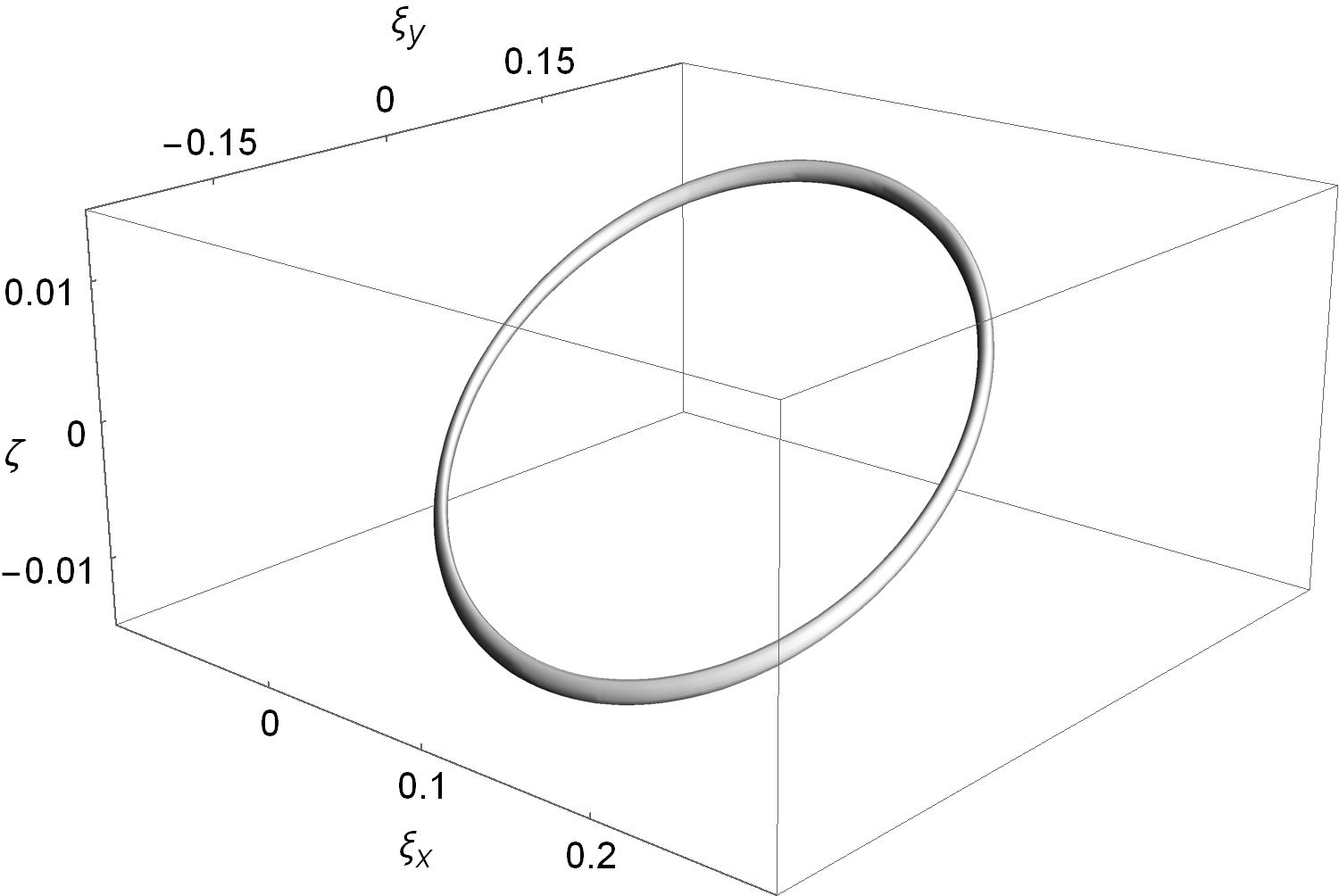}
\end{center}
\vspace*{-4.5ex}
\caption{The nodal line of $\Psi(\bm{\xi},\zeta)$ given by the formula~(\ref{fullfring}). The distances on the axes are measured in $k^{-1}=\lambda/2\pi$. The values of parameters are as follows: $\gamma=10$, $\kappa=0.01$, $\chi=0.01$, $\chi_1= 0.009$ and $\chi_2 = 0.01$.}
\label{besselring}
\end{figure}

Applying~(\ref{evol}) together with~(\ref{propagator}) one obtains the paraxial envelope in the form
\begin{eqnarray}
\Psi(\bm{\xi},\zeta) =&& \frac{1}{2\pi}\sum_{n=-\infty}^{\infty}\int\limits_0^\infty dt\, t\int\limits_0^\infty d\xi' \xi'\int\limits_0^{2\pi} d\phi' e^{in(\phi-\phi')}e^{-\frac{i\zeta t^2}{2}}\nonumber\\
&&\times J_n(t\xi)J_n(t\xi')\big[\alpha_1J_0(\chi_1\xi')+\alpha_2J_0(\chi_2\xi')\nonumber\\
&&+\beta e^{i\phi'} J_1(\chi\xi')\big]e^{-\kappa \xi'^2},\label{frev}
\end{eqnarray}
Integral with respect to $\phi'$ can be easily taken and leads to the two Kronecker deltas: $\delta_{n0}$ (the first two terms) and $\delta_{n1}$ (the last one), which reduces the infinite sum over $n$ to two terms only:
\begin{eqnarray}
\Psi(\bm{\xi},\zeta) =&&\int\limits_0^\infty dt\, te^{-\frac{i\zeta t^2}{2}} \int\limits_0^\infty d\xi' \xi' e^{-\kappa \xi'^2}\big[J_0(t\xi)J_0(t\xi')\label{frev1}\\
&&\times\sum_{l=1}^2\alpha_lJ_0(\chi_l\xi')+\beta e^{i\phi} J_1(t\xi)J_1(t\xi')J_1(\chi\xi')\big],\nonumber
\end{eqnarray}
The remaining integrals with respect to $\xi'$ and $t$ are of similar character and can be executed subsequently with the use of the formulas~\cite{gr}:
\begin{subequations}\label{bes}
\begin{align}
\int\limits_0^\infty dx\, x\, e^{- p x^2}J_n(ax)J_n(bx)=&\frac{1}{2p}\,e^{-\frac{a^2+b^2}{4p}}I_n\Big(\frac{ab}{2p}\Big),\nonumber\\
&\mathrm{for}\;\; p\in\mathbb{R}_+\label{bes1}\\
\int\limits_0^\infty dx\, x\, e^{- q x^2}J_n(ax)I_n(bx)=&\frac{1}{2q}\,e^{-\frac{a^2-b^2}{4q}}J_n\Big(\frac{ab}{2q}\Big),\nonumber\\
&\mathrm{for}\;\;\mathrm{Re}\, q\in\mathbb{R}_+\label{bes2}
\end{align}
\end{subequations}
As a result of these two integrations, the complete paraxial envelope, applicable for $\zeta\neq 0$, is found as a sum of three coaxial BG modes: 
\begin{eqnarray}
\Psi(\bm{\xi},\zeta) =&& \frac{1}{c(\zeta)}e^{-\frac{\kappa\xi^2}{c(\zeta)}}\bigg[\sum_{l=1}^2\alpha_le^{-i\frac{\chi_l^2\zeta}{2c(\zeta)}}J_0\Big(\frac{\chi_l\xi}{c(\zeta)}\Big)\nonumber\\
&&+\beta e^{i\phi}e^{-i\frac{\chi^2\zeta}{2c(\zeta)}} J_1\Big(\frac{\chi\xi}{c(\zeta)}\Big)\bigg],
\label{fullfring}
\end{eqnarray}
where $c(\zeta)=1+2i \kappa\zeta$. 
The nodal line of $\Psi(\bm{\xi},\zeta)$, can now be easily drawn. It is done in Fig.~\ref{besselring}. As expected, it represents a ring. What should be especially emphasized, it is entirely constructed of physical beams of BG type.

The form of $\Psi(\bm{\xi},\zeta)$ as given in~(\ref{fullfring}) may be somewhat illegible to an optical physicist due to the notation used. This notation is highly convenient for conducting calculations, as many of the expressions greatly simplify, but below the result is rewritten in a traditional form in which the standard coaxial BG modes can be easily recognized. 
The connection will be easily established if one observes that
\begin{subequations}\label{nota}
\begin{align}
c(\zeta)&=1+i\frac{z}{z_R},\;\;\;\; \frac{1}{c(\zeta)}=\frac{w_0}{w(z)}e^{-i\psi(z)},\label{nota1}\\
\chi_l &=\sin\theta_l,\;\;\;\; \chi =\sin\theta \label{nota2}\\
\kappa&=\frac{1}{k^2w_0^2},\label{nota3}
\end{align}
\end{subequations}
where $z_R$ denotes the Rayleigh length, $w_0$ is the beam waist, $w(z)=w_0\sqrt{1+(z/z_R)^2}$ stands for the beam radius, $R(z)=z(1+(z_R/z)^2)$ denotes the wave-front curvature, $\psi(z)=\arctan (z/z_R)$ is the Gouy phase and $\theta_l$ (and similarly $\theta$) denotes the angular half-aperture of the appropriate beam cone. 
For small values of these latter quantities expressed in radians, just as it is in this work, they are practically equal to the parameters $\chi$.

As can be seen from~(\ref{fullfring}) and with the application of the definitions~(\ref{nota}), the desired nodal line of Fig.~\ref{besselring} can be constructed as a result of the superposition of three standard BG beams, readily obtainable in experiments, of the form:
\begin{eqnarray}
&&\Psi_n(\bm{r},z)=\frac{w_0}{w(z)}e^{in\phi}J_n\Big(\frac{kr\sin\theta}{1+iz/z_R}\Big)\label{super}\\
&&\;\;\;\;\;\;\exp\left[-\frac{r^2}{w(z)^2}-i\frac{kr^2 }{2 R(z)}-i\frac{z/z_R}{1+iz/z_R}\sin^2\theta-i\psi(z)\right],\nonumber
\end{eqnarray}
with $n=0$ or $n=1$ and different values of the aperture angles $\theta_1,\,\theta_2$ and $\theta$, and with intensities determined by the coefficients $\alpha_l$ and $\beta$. Since only relative values matter, $\alpha_1$ may be by definition set to $1$ and then, for the data of Fig.~\ref{besselring} one gets 
$\alpha_2=-1,\, \beta\approx -0.00019$. It is interesting to note that the $\phi$-dependent contribution is much weaker than the principal ones containing Bessel functions of the zeroth order. They are, however, necessary to ensure the correct shape of the nodal line, close to which the contributions from the ``large'' terms strongly decrease. This means that creating a knotted line through the mechanism of the destructive interference is a delicate matter requiring some precision and the nodal line itself is a rather shallow structure. This observation is confirmed by our further examples. The values of the wavelength, beam waist and Rayleigh length are identical for all three beams and can be adapted to the experimental requirements.

\begin{figure}[h]
\begin{center}
\includegraphics[width=0.45\textwidth,angle=0]{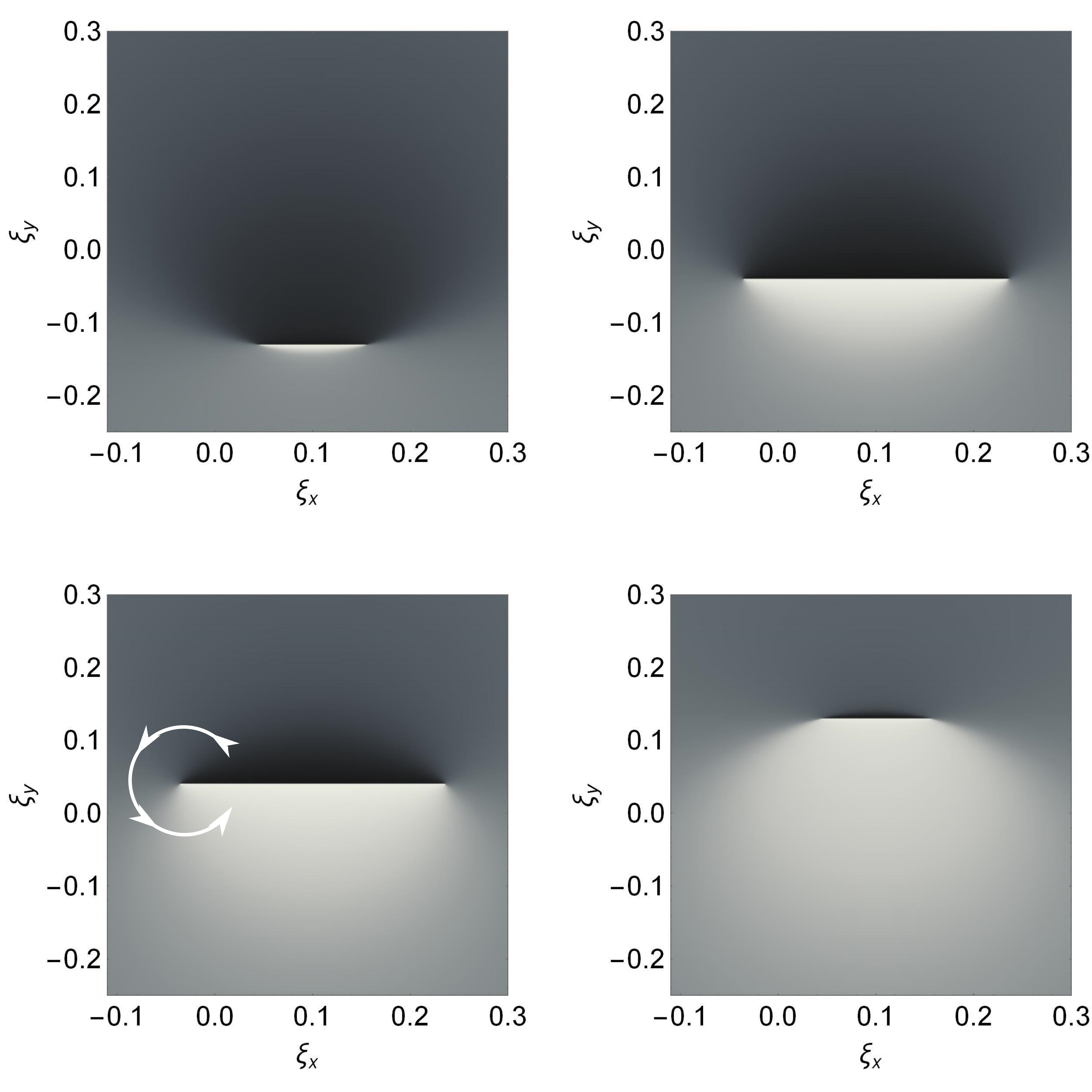}
\end{center}
\vspace*{-4.5ex}
\caption{The phases of the light beam created as a superposition (\ref{fullfring}) in four planes: $\zeta=-0.013,\; -0.004,\; 0.004,\; 0.013$. The value of the phase is represented continuously by the means of the grayscale from $-\pi$ (black color) to $\pi$ (white color).}
\label{besselringps}
\end{figure}

In Fig. \ref{besselringps} the phases of the outgoing light beam are depicted in four selected planes $\zeta=\mathrm{const}$. In particular, one can see the phase change from the value of $-\pi$ (black color) to the value of $\pi$ (white color) when walking around the nodal line (which is -- roughly speaking --perpendicular to the plane at that point) as shown in the figure. This is typical for vortex lines from which a knot is formed. The same observation can be made in the following figures.

\subsection{The Hopf link}\label{hopf}

The {\em Hopf link} is the first nontrivial knot and represents two rings linked to each other. It can again be obtained from~(\ref{qgen}) upon setting $n=2$:
\begin{equation}
q(u,v)=(u-v)(u+v).
\label{qhopf}
\end{equation}
The corresponding Milnor polynomial, obtained by substituting $u$ and $v$ as defined by~(\ref{stereo}), and next reduced to the plane $\zeta=0$, takes the form
\begin{equation}
q_M(\bm{\xi},0)=(1-\xi_x^2-\xi_y^2)^2-4(\xi_x+i\xi_y)^2.
\label{fhopf}
\end{equation}
and that with the scaling factor
\begin{equation}
q_M(\bm{\xi},0)=[1-\gamma^2(\xi_x^2+\xi_y^2)]^2-4\gamma^2(\xi_x+i\xi_y)^2.
\label{fhopfg}
\end{equation}
Analyzing the presence of the factors $e^{i m\phi}$ in the expression above (where $\xi_x+i\xi_y\mapsto \xi e^{i\phi}$) one easily notices, that now Bessel functions of two orders ($m=0$ and $m=2$) are required. On the other hand even for $J_0$ one needs to fix terms up to $\xi^4$. Since $J_0$ in an even function, this means that the sum with respect to $l$ in~(\ref{suma}) now extends to $l=3$. Therefore, one can replace $q_M(\bm{\xi},0)$ with 
\begin{eqnarray}
q_M(\bm{\xi},0)\mapsto \sum_{l=1}^3\alpha_l J_0(\chi_l\xi)+\beta e^{2i\phi} J_2(\chi\xi)
\label{intb}
\end{eqnarray}
The values of coefficients are established similarly as it was done in the former subsection. They turn out to be 
\begin{equation}
\alpha_l=\prod_{j=1\atop j\neq l}^3 \frac{\chi_j^2-8\gamma^2}{\chi_l^2-\chi_j^2},\;\;\;\;  \beta=-\frac{32\gamma^2}{\chi^2}.
\label{aabi}
\end{equation}

\begin{figure}[b]
\begin{center}
\includegraphics[width=0.45\textwidth,angle=0]{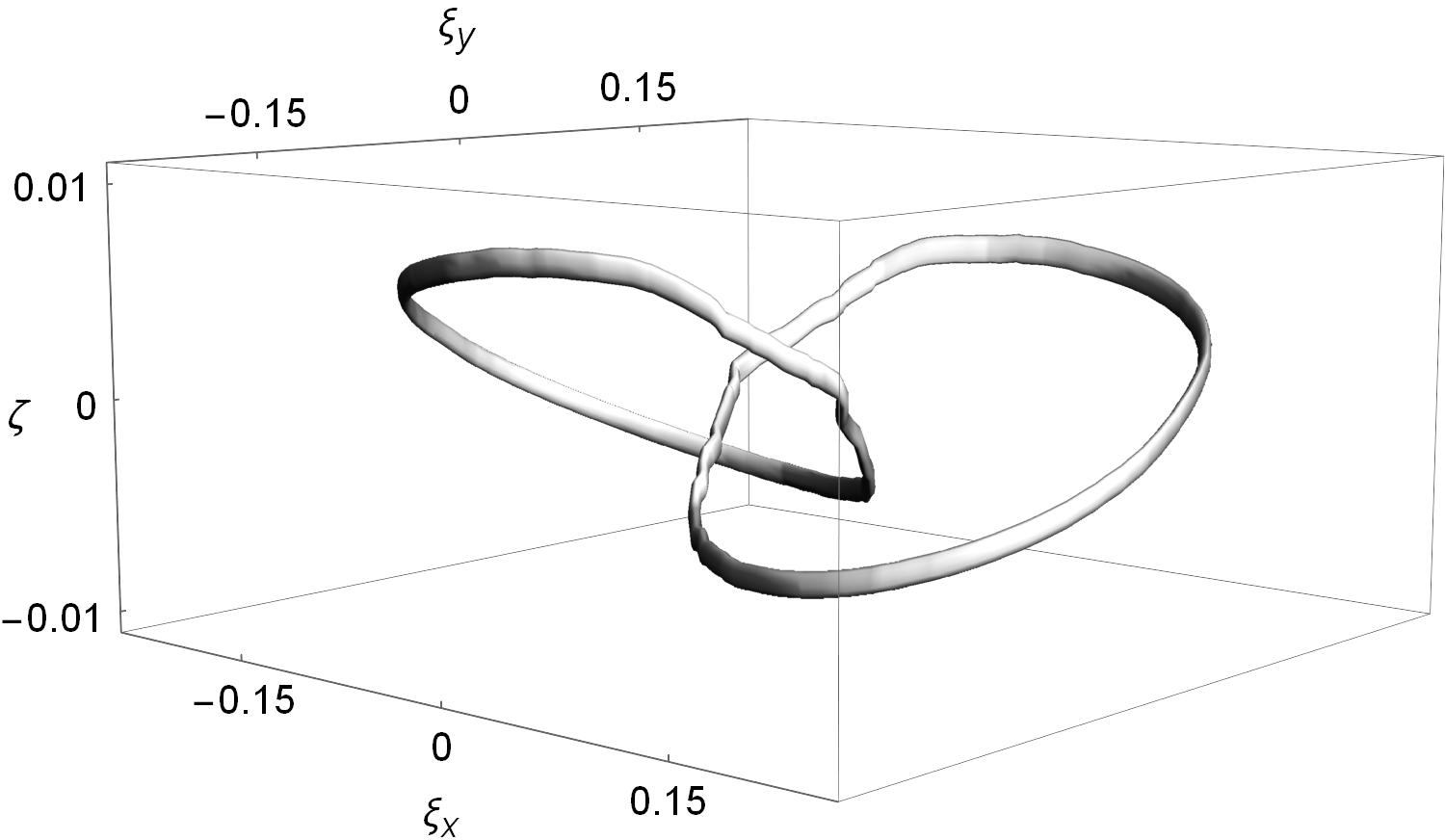}
\end{center}
\vspace*{-4.5ex}
\caption{The nodal lines of $\Psi(\bm{\xi},\zeta)$ given by the formula~(\ref{fullhopf}). Units on axes as well as values of $\gamma$ and $\kappa$ are identical as in Fig.~\ref{besselring}. The values of other parameters are as follows: $\chi=0.01$, $\chi_1= 0.008$, $\chi_2 = 0.009$ and $\chi_3 = 0.01$.}
\label{besselhopf}
\end{figure}

Following the procedure outlined in the case of the unknot, one now finds
\begin{eqnarray}
&&\Psi(\bm{\xi},\zeta) =\frac{1}{2\pi}\sum_{n=-\infty}^{\infty}\int\limits_0^\infty dt\, t\int\limits_0^\infty d\xi' \xi'\int\limits_0^{2\pi} d\phi' e^{in(\phi-\phi')}e^{-\frac{i\zeta t^2}{2}}\nonumber\\
&&\;\;\;\;\times J_n(t\xi)J_n(t\xi')\bigg[\sum_{l=1}^3\alpha_lJ_0(\chi_l\xi')+\beta e^{2i\phi'} J_2(\chi\xi')\bigg]e^{-\kappa \xi'^2}.\nonumber\\
\label{fhev}
\end{eqnarray}
Performing the trivial integration over $\phi'$ together with the $n$ summation and next those with respect to $\xi'$ and $t$ according to~(\ref{bes}), one finally finds
\begin{eqnarray}
\Psi(\bm{\xi},\zeta) =&& \frac{1}{c(\zeta)}e^{-\frac{\kappa\xi^2}{c(\zeta)}}\bigg[\sum_{l=1}^3\alpha_le^{-i\frac{\chi_l^2\zeta}{2c(\zeta)}}J_0\Big(\frac{\chi_l\xi}{c(\zeta)}\Big)\nonumber\\
&&+\beta e^{2i\phi}e^{-i\frac{\chi^2\zeta}{2c(\zeta)}} J_2\Big(\frac{\chi\xi}{c(\zeta)}\Big)\bigg].
\label{fullhopf}
\end{eqnarray}
The obtained envelope turns out to be a superposition of four BG modes defined with~(\ref{super}) with $n=0$ and $n=2$. The nodal lines are depicted in Fig.~\ref{besselhopf} and clearly constitute the Hopf link.

The normalized (i.e. after having set $\alpha_1=1$) values of coefficients are: $\alpha_2\approx -1.89,\, \alpha_3\approx 0.89,\,\beta\approx-2.82\times 10^{-8}$, and allow to establish the relative intensity of the combined beams. Again, a very tiny value of the coefficient of the angle-dependent term is worth noting. Obviously, it can be modified within certain limits by a suitable choice of $\chi$'s, but this coefficient is invariably significantly smaller. In turn, the negative values account for the relative phases of the modes.

It is noteworthy that the knot lines produced by this method turn out to be pretty smooth (cf. e.g.~\cite{bkj}), which may have practical significance.

\begin{figure}[h]
\begin{center}
\includegraphics[width=0.45\textwidth,angle=0]{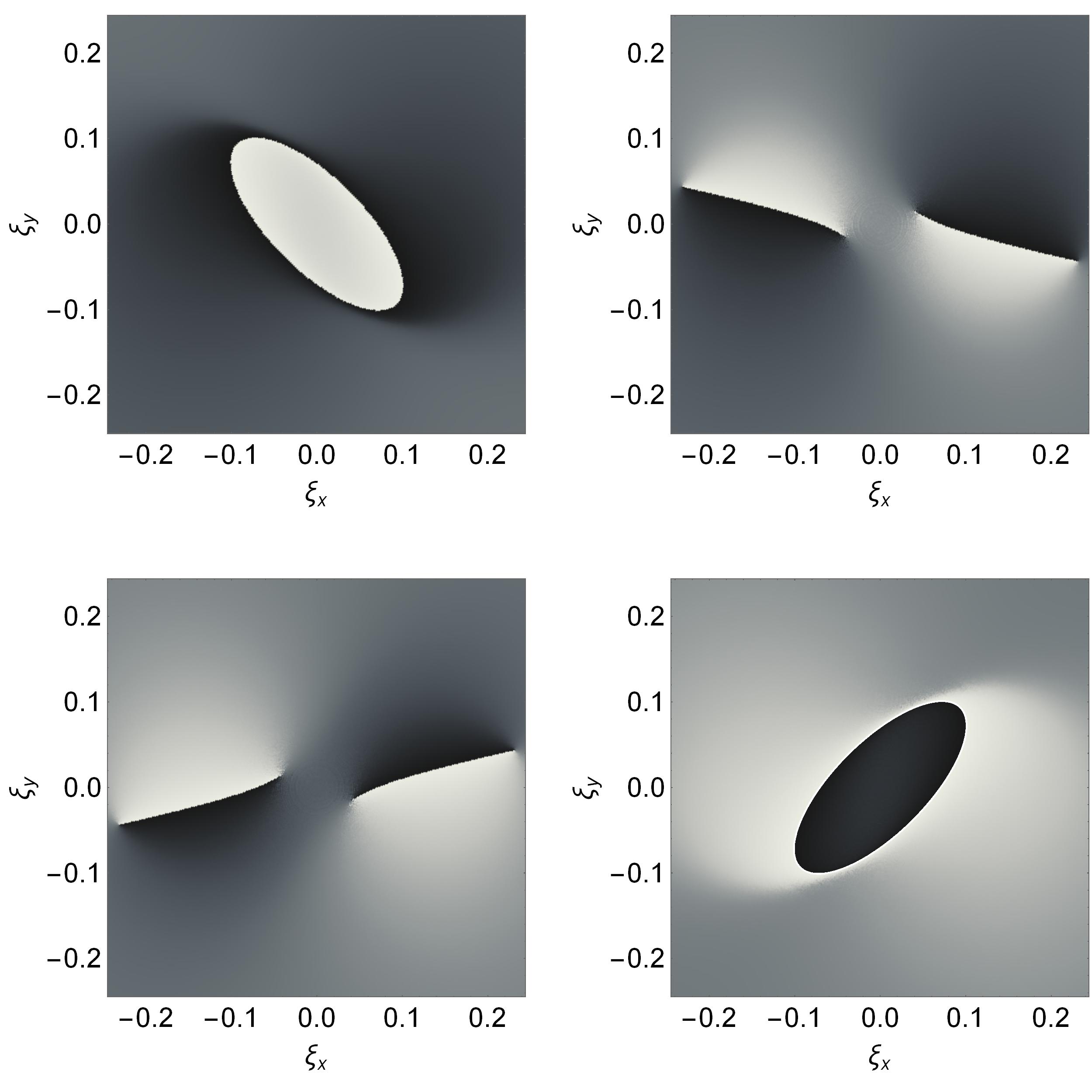}
\end{center}
\vspace*{-4.5ex}
\caption{The phases of the light beam created as a superposition (\ref{fullhopf}) in four planes: $\zeta=-0.007,\; -0.002,\; 0.002,\; 0.007$.}
\label{besselhopfps}
\end{figure}

The phases of $\Psi(\bm{\xi},\zeta)$ are drawn in Fig. \ref{besselhopfps}. Again the change of the phase by $2\pi$ can be observed, when encircling the vortex line.

\subsection{The Borromean rings}\label{borings}

Proceeding further towards the higher complexity of considered knots, we will now examine that composed of three loops, each of which is linked to both others. Such a knotted construction bears the name of {\em the Borromean rings}, and is again obtainable from~(\ref{qgen}) when setting $n=3$:
\begin{equation}
q(u,v)=(u-v)(u-e^{2\pi i/3}v)(u-e^{4\pi i/3}v).
\label{qbor}
\end{equation}
In consequence, upon applying the outlined procedure the corresponding Milnor polynomial if found to be
\begin{equation}
q_M(\bm{\xi},0)=(-1+\gamma^2(\xi_x^2+\xi_y^2))^3-8\gamma^3(\xi_x+i\xi_y)^3,
\label{fbor}
\end{equation}
where the scaling factor has already been introduced. Upon the expansion, this expression contains the following powers of $\xi$: $ \xi^0, \xi^2, \xi^4$ and $\xi^6$, as well as the term with the factor $e^{3i\phi}$. This indicates that four zero-order Bessel functions together with $J_3$ will be needed:
\begin{equation}
q_M(\bm{\xi},0)\mapsto \sum_{l=1}^4\alpha_l J_0(\chi_l\xi)+\beta e^{3i\phi} J_3(\chi\xi),
\label{intc}
\end{equation}
where coefficients can be calculated according to the following formulas:
\begin{eqnarray}
\alpha_l= &&\frac{\prod_{j=1\atop j\neq l}^4(\chi_j^2-12\gamma^2)+48\gamma^4(\sum_{j=1\atop j\neq l}^4 \chi_j^2-12\gamma^2)}{\prod_{j=1\atop j\neq l}^4 (\chi_l^2-\chi_j^2)},\nonumber\\
\beta=&&  -\frac{384\gamma^3}{\chi^3}.\label{abn}
\end{eqnarray}

This allows to write the integral form of the enevelope:
\begin{eqnarray}
&&\Psi(\bm{\xi},\zeta) =\frac{1}{2\pi}\sum_{n=-\infty}^{\infty}\int\limits_0^\infty dt\, t\int\limits_0^\infty d\xi' \xi'\int\limits_0^{2\pi} d\phi' e^{in(\phi-\phi')}e^{-\frac{i\zeta t^2}{2}}\nonumber\\
&&\;\;\;\;\times J_n(t\xi)J_n(t\xi')\bigg[\sum_{l=1}^4\alpha_lJ_0(\chi_l\xi')+\beta e^{3i\phi'} J_3(\chi\xi')\bigg]e^{-\kappa \xi'^2}.\nonumber\\
\label{fbev}
\end{eqnarray}

\begin{figure}[b]
\begin{center}
\includegraphics[width=0.45\textwidth,angle=0]{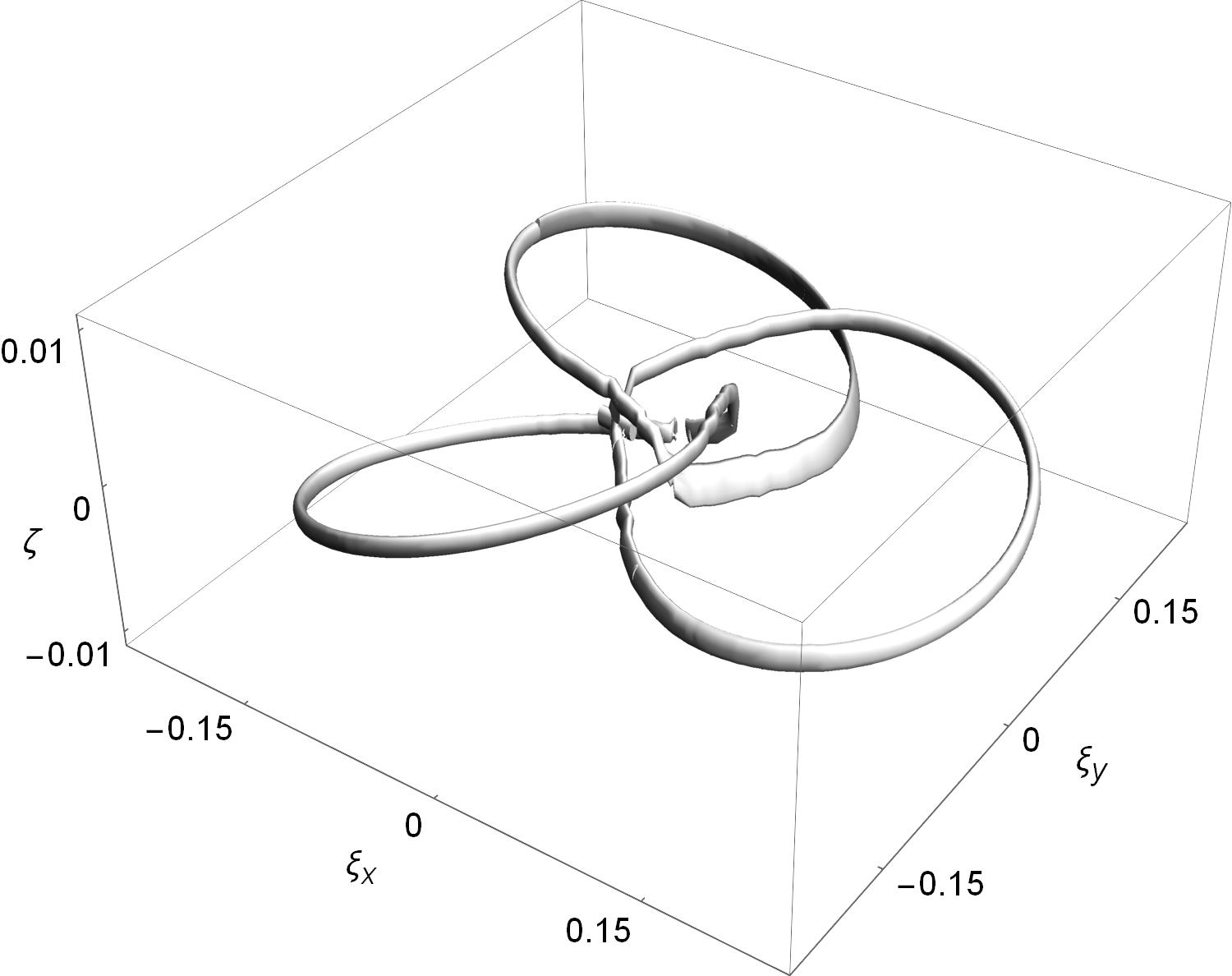}
\end{center}
\vspace*{-4.5ex}
\caption{Same as in Fig. \ref{besselring} but for the nodal lines of $\Psi(\bm{\xi},\zeta)$ given by the formula~(\ref{fullbevf}). The values of parameters are as follows: $\chi=0.001$, $\chi_1= 0.05$, $\chi_2 = 0.1$, $\chi_3 = 0.15$ and $\chi_4 = 0.2$.}
\label{besselbors}
\end{figure}

Executing the integrals and the sum as in the previous cases, one arrives at
\begin{eqnarray}
\Psi(\bm{\xi},\zeta) =&& \frac{1}{c(\zeta)}e^{-\frac{\kappa\xi^2}{c(\zeta)}}\bigg[\sum_{l=1}^4\alpha_le^{-i\frac{\chi_l^2\zeta}{2c(\zeta)}}J_0\Big(\frac{\chi_l\xi}{c(\zeta)}\Big)\nonumber\\
&&+\beta e^{3i\phi}e^{-i\frac{\chi^2\zeta}{2c(\zeta)}} J_3\Big(\frac{\chi\xi}{c(\zeta)}\Big)\bigg]
\label{fullbevf}
\end{eqnarray}

The nodal lines of $\Psi(\bm{\xi},\zeta)$ are demonstrated in Fig. \ref{besselbors}. Obviously, they represent a system of three linked loops, i.e. the Boromean rings. The superposition yielding such a rather sophisticated knot consists of five BG mods (\ref{fullbevf}), each defined by the formula (\ref{super}).

It is quite a challenge to produce these lines because in places of their close passing, the calculations require great precision. It is related to the earlier observation that the factor $\beta$ is much smaller than $\alpha_l$'s, which applies in the present case. This means again that generating such a structure needs a precise adjustment of the intensity of the beam bearing orbital angular momentum. For better visualization the values of $\chi_{1,2,3,4}$ in Fig. \ref{besselbors} have been increased and that of $\chi$ decreased. This makes all coefficients comparable in size and the graphic representation is easier to obtain in a numerical manner. Large differences between the coefficient values, reaching many orders of magnitude, require very high calculation accuracy. 

The normalized values of the coefficients are now as follows:
\begin{subequations}\label{val}
\begin{align}
&\alpha_2\approx -2.00001,\;\;\;\; \alpha_3\approx 1.28574,\label{val1}\\
&\alpha_4\approx -0.285723,\;\;\;\; \beta\approx -0.937557.\label{val2}
\end{align}
\end{subequations}
They are recorded with the accuracy of a few significant digits to illustrate how crucial is the precision for obtaining a particular knot. This is related to our discussion concerning the Boromean rings, so we will reconsider these values later.

When decreasing values of $\chi_i$'s in the beams, $\beta$ (to put it precisely, the normalized value of $\beta$, i.e. $\beta/\alpha_1$) becomes extremely small, as in previous cases which reflects the complicated nodal structure close to the passing points.

\begin{figure}[h]
\begin{center}
\includegraphics[width=0.45\textwidth,angle=0]{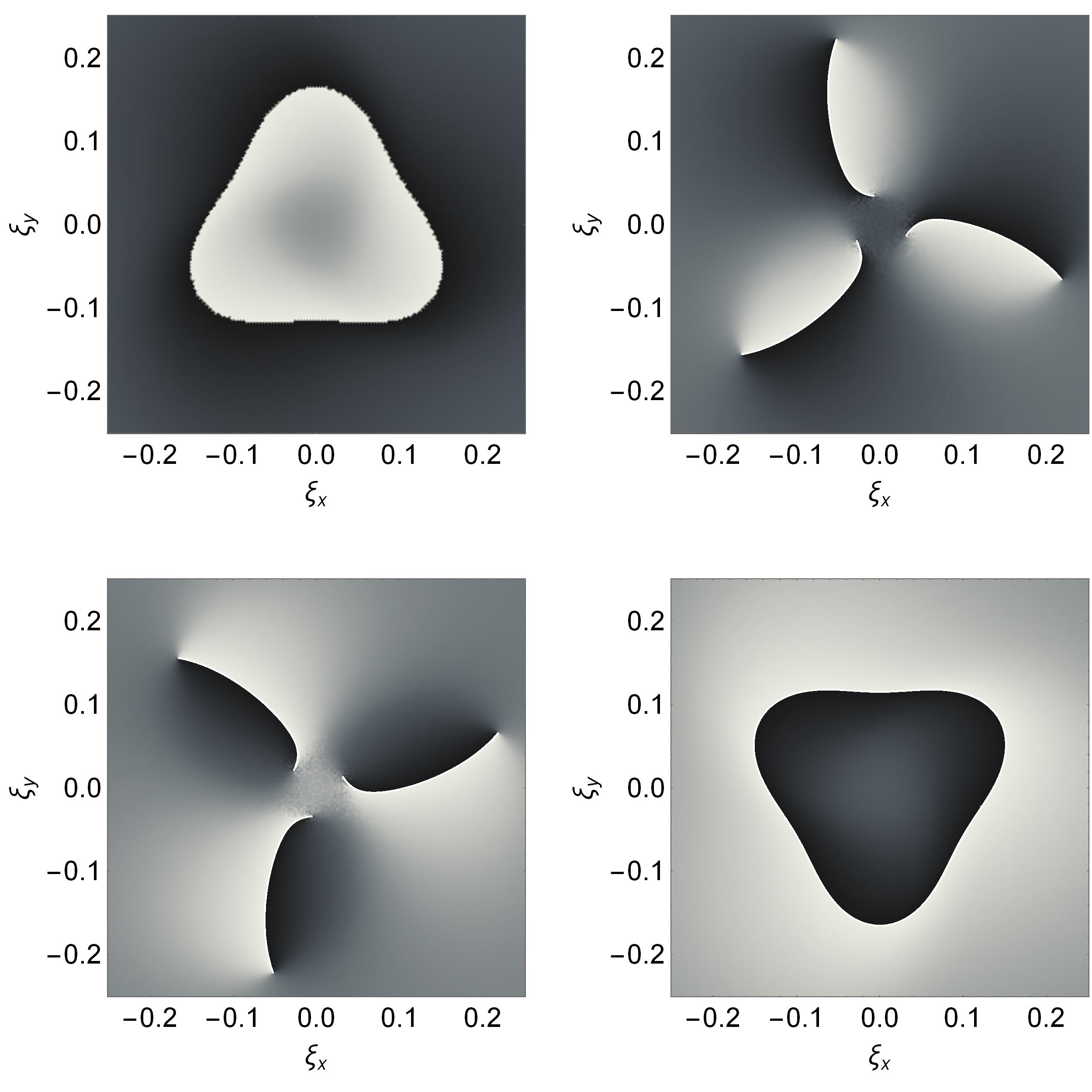}
\end{center}
\vspace*{-4.5ex}
\caption{Same as in Fig. \ref{besselhopfps}, but for formula (\ref{fullbevf}).}
\label{besselborps}
\end{figure}

In Fig. \ref{besselborps} the phases of  (\ref{fullbevf}) are drawn in four planes. The same effects as before are observed.

\subsection{The trefoil}\label{trefoil}

In order to obtain the knot called {\em the trefoil} one has to start with the polynomial
\begin{equation}
q(u,v)=u^2-v^3.
\label{qtre}
\end{equation}
which does not belong to the class described with the formula~(\ref{qgen}) but falls into a wide family of knots that can be derived from the expressions of the type
\begin{equation}
q(u,v)=u^2-v^n, 
\label{qtren}
\end{equation}
with $n\in \mathbb{N}$~\cite{king}. The same family includes for example the cinquefoil knot (for $n=5$) or the septafoil knot (for $n=7$) etc. Setting $n=3$ the following Milnor polynomial (with the scaling factor $\gamma$ included) is found:
\begin{equation}
q_M(\bm{\xi},0)=[1+\gamma^2(\xi_x^2+\xi_y^2)][1-\gamma^2(\xi_x^2+\xi_y^2)]^2-8\gamma^3(\xi_x+i\xi_y)^3,
\label{ftre}
\end{equation}

\begin{figure}[b]
\begin{center}
\includegraphics[width=0.45\textwidth,angle=0]{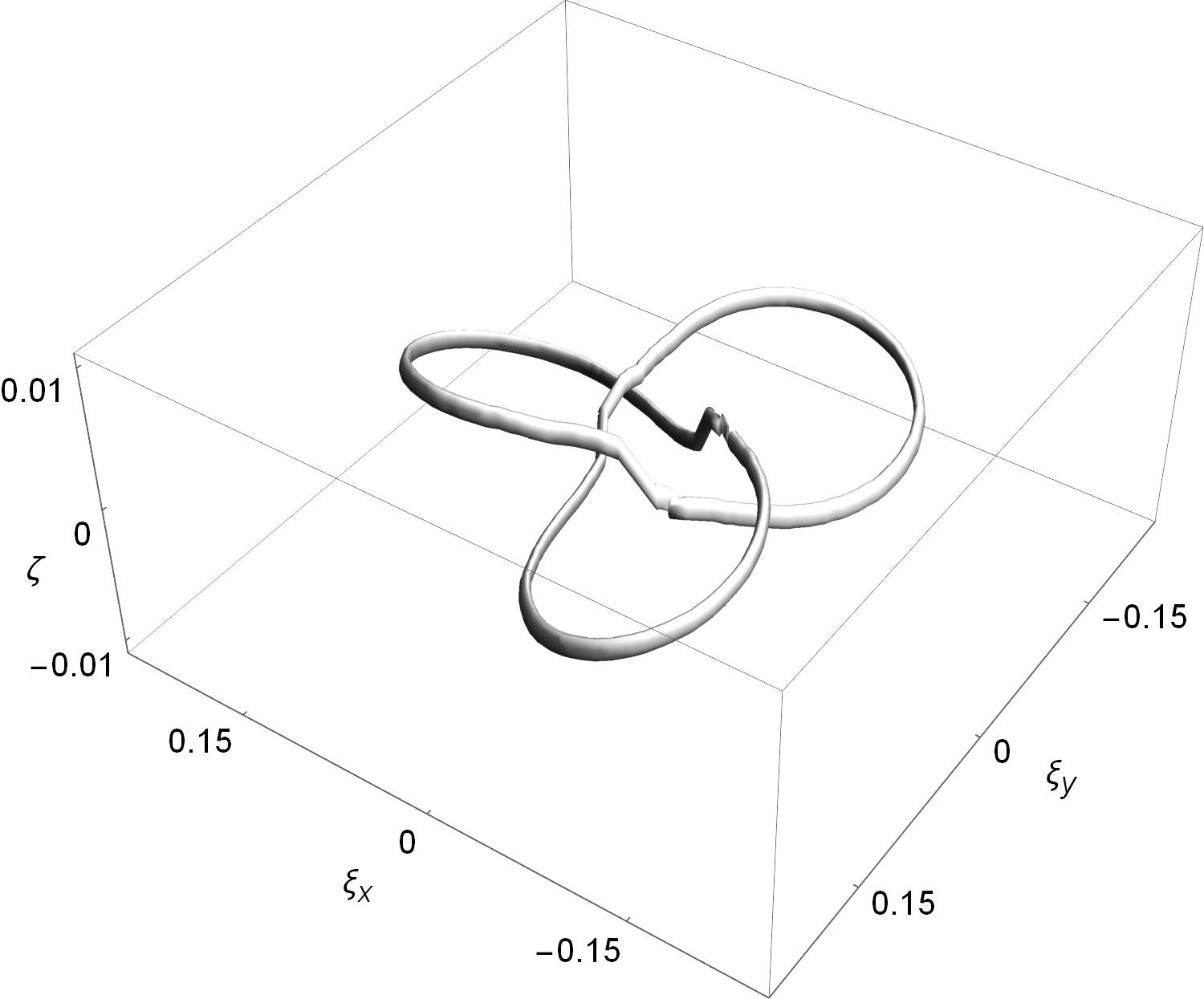}
\end{center}
\vspace*{-4.5ex}
\caption{Same as in Fig. \ref{besselbors} but for the nodal lines of $\Psi(\bm{\xi},\zeta)$ given by the formula~(\ref{fullbevf}) with the coefficients~(\ref{abtre}). All the values of parameters are identical.}
\label{besseltre}
\end{figure}

The wave envelope can now be constructed from $q_M(\bm{\xi},0)$ following the steps of the previous subsections. Again the expression~(\ref{fullbevf}) is obtained, since $q_M(\bm{\xi},0)$ contains the same terms (apart from the multiplicative constants) as in the case of the Borromean rings. Only the values of the coefficients alter (i.e. the relative intensities of the beams involved) and they are currently expressed as follows 
\begin{eqnarray}
\alpha_l= &&-\frac{\prod_{j=1\atop j\neq l}^4(\chi_j^2-4\gamma^2)+48\gamma^4(\sum_{j=1\atop j\neq l}^4 \chi_j^2+48\gamma^2)+64\gamma^6}{\prod_{j=1\atop j\neq l}^4 (\chi_l^2-\chi_j^2)},\nonumber\\
\beta=&& -\frac{384\gamma^3}{\chi^3},\label{abtre}
\end{eqnarray}
where minuses are obviously inessential.

The nodal line obtained now from~(\ref{fullbevf}) has really the form of a trefoil and is depicted in Fig.~\ref{besseltre}. The values of parameters (i.e. the beams used) are identical as in the case of the Borromean rings. Only their relative intensities differ. This means that by setting up experimentally an identical set of BG beams, various knots of entirely different topology can be obtained by simply modifying their relative intensities. Formulas~(\ref{abtre}) yield now the following values of the ratios:  
\begin{subequations}\label{vala}
\begin{align}
&\alpha_2\approx -1.99999583,\;\;\;\; \alpha_3\approx 1.28571,\label{vala1}\\
&\alpha_4\approx -0.285711,\;\;\;\; \beta\approx -0.937481.\label{vala2}
\end{align}
\end{subequations}
They should be compared to (\ref{val}). The differences between these values apparently seem negligible, but the topology of the knots is completely distinct. Tiny changes in the relative intensities of the beams may lead to a different reconnection of nodal lines passing very close to one another and to a completely different knot. This property seems to be general: for any knot with a more complex structure, there may happen a kind of ``switching'' of nodal lines.

\begin{figure}[h]
\begin{center}
\includegraphics[width=0.45\textwidth,angle=0]{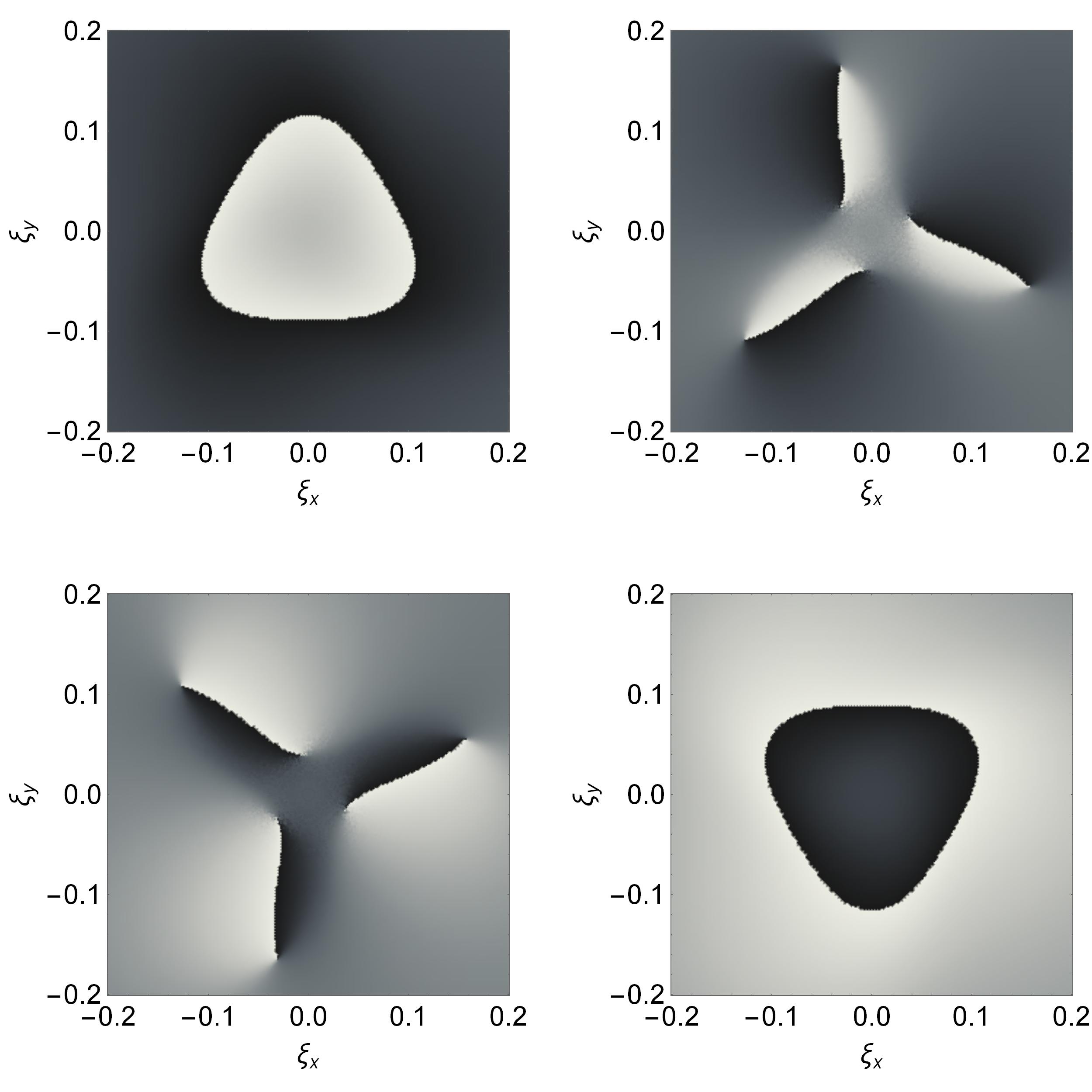}
\end{center}
\vspace*{-4.5ex}
\caption{Phase reduced to -pi pi}
\label{besseltrelps}
\end{figure}

Fig. \ref{besseltrelps} presents the phases in the planes of constant $\zeta$ for the trefoil knot. The similarities to the Borromean rings case should be noted.

\subsection{More complicated knots}\label{cok}

We now move on to more complex knots which are simultaneously more challenging to generate numerically and experimentally using BG light beams. As an example let us consider {\em the figure-eight knot} as its shape -- if open, as usually is in climbing or sailing -- with the appropriate line arrangement in space, resembles number $8$. Contrary to that, the mathematical knot is closed, according to the definition provided in Section~\ref{genpro}. The polynomial $q(u,v)$ is in this case considerably more complicated, and in addition to the complex variables $u$ and $v$ depends as well on the conjugated quantity $\bar{v}$~\cite{bkj,bode}:
\begin{equation}
q(u,v)=64u^3-12u[2(v^2-\bar{v}^2)+3]-14(v^2+\bar{v}^2)-(v^4-\bar{v}^4)
\label{qf8}
\end{equation}
The Minor polynomial contains now many terms, which means that in order to create the beam of this topology, relatively many BG modes have to be superimposed. When rewritten in polar variables it contains expressions up to $\xi^8$ and also angular factors $e^\pm 2i\phi$ and $e^\pm 4i\phi$:
\begin{eqnarray}
q_M(\bm{\xi},0)&=&28[\gamma^8(\xi_x^2+\xi_y^2)^4-1]\nonumber\\
&& -200\gamma^2(\xi_x^2+\xi_y^2)[\gamma^4(\xi_x^2+\xi_y^2)^2-1]\nonumber\\
&&+\gamma^2(\xi_x^2+\xi_y^2)[-152\gamma^4(\xi_x^2+\xi_y^2)^2\nonumber\\
&&-112\gamma^2(\xi_x^2+\xi_y^2)+40]e^{2i\phi}\label{f8}\\
&&+\gamma^2(\xi_x^2+\xi_y^2)[40\gamma^4(\xi_x^2+\xi_y^2)^2\nonumber\\
&&-112\gamma^2(\xi_x^2+\xi_y^2)-152]e^{-2i\phi}\nonumber\\
&&-16\gamma^4(\xi_x^2+\xi_y^2)^2e^{4i\phi}+16\gamma^4(\xi_x^2+\xi_y^2)^2e^{-4i\phi}.\nonumber
\end{eqnarray}
It may be checked by the Taylor expansion of Bessel functions that the field envelope has to be written the form:
\begin{eqnarray}
&&\Psi(\bm{\xi},\zeta) =\frac{1}{2\pi}\sum_{n=-\infty}^{\infty}\int\limits_0^\infty dt\, t\int\limits_0^\infty d\xi' \xi'\int\limits_0^{2\pi} d\phi' e^{in(\phi-\phi')}e^{-\frac{i\zeta t^2}{2}}\nonumber\\
&&\;\;\;\;\times J_n(t\xi)J_n(t\xi')\bigg[\sum_{l=1}^5\alpha_lJ_0(\chi_l\xi')+\sum_{l=1}^3\beta_l e^{2i\phi'} J_2(\chi_l'\xi')\nonumber\\
&&\;\;\;\;+\sum_{l=1}^3\gamma_l e^{-2i\phi'} J_{-2}(\chi_l''\xi')+\delta_1 e^{4i\phi'} J_4(\chi\xi')\nonumber\\
&&\;\;\;\;+\delta_2 e^{-4i\phi'} J_{-4}(\chi'\xi')\bigg]e^{-\kappa \xi'^2},
\label{ff8ev}
\end{eqnarray}
where the coefficients have to satisfy the sets of equations listed below. The first one for $\alpha_l$'s comprises the following five equations: 
\begin{subequations}\label{alphas}
\begin{align}
\alpha_1+\alpha_2+\alpha_3+\alpha_4+\alpha_5&=28,\label{alphas1}\\
\alpha_1\chi_1^2+\alpha_2\chi_2^2+\alpha_3\chi_3^2+\alpha_4\chi_4^2+\alpha_5\chi_5^2&=800\gamma^2,\label{alphas2}\\
\alpha_1\chi_1^4+\alpha_2\chi_2^4+\alpha_3\chi_3^4+\alpha_4\chi_4^4+\alpha_5\chi_5^4&=0,\label{alphas3}\\
\alpha_1\chi_1^6+\alpha_2\chi_2^6+\alpha_3\chi_3^6+\alpha_4\chi_4^6+\alpha_5\chi_5^6&=-460800\gamma^6,\label{alphas4}\\
\alpha_1\chi_1^8+\alpha_2\chi_2^8+\alpha_3\chi_3^8+\alpha_4\chi_4^8+\alpha_5\chi_5^8&=-4128768\gamma^8.\label{alphas5}
\end{align}
\end{subequations}
That for $\beta_l$'s is composed of three equations:
\begin{subequations}\label{betas}
\begin{align}
\beta_1{\chi'}_1^2+\beta_2{\chi'}_2^2+\beta_3{\chi'}_3^2&=320\gamma^2,\label{betas1}\\
\beta_1{\chi'}_1^4+\beta_2{\chi'}_2^4+\beta_3{\chi'}_3^4&=10752\gamma^4,\label{betas2}\\
\beta_1{\chi'}_1^6+\beta_2{\chi'}_2^6+\beta_3{\chi'}_3^6&=-466944\gamma^6,\label{betas3}
\end{align}
\end{subequations}
 and is similar to the set for $\gamma_l$'s:
\begin{subequations}\label{gammas}
\begin{align}
\gamma_1{\chi''}_1^2+\gamma_2{\chi''}_2^2+\gamma_3{\chi''}_3^2&=-1216\gamma^2,\label{gammas1}\\
\gamma_1{\chi''}_1^4+\gamma_2{\chi''}_2^4+\gamma_3{\chi''}_3^4&=10752\gamma^4,\label{gammas2}\\
\gamma_1{\chi''}_1^6+\gamma_2{\chi''}_2^6+\gamma_3{\chi''}_3^6&=122880\gamma^6,\label{gammas3}
\end{align}
\end{subequations}
Solutions to all of these equations can be found in an obvious way. We do not write them out explicitly in order to avoid listing lenghty expressions. Unlike, the last two parameters can directly be found:
\begin{equation}
\delta_1=-\frac{6144\gamma^4}{\chi^4},\;\;\;\; \delta_2=\frac{6144\gamma^4}{{\chi'}^4}
\label{deltas.}
\end{equation}

Now, we are in the position to present the result for the envelope:
\begin{eqnarray}
\Psi(\bm{\xi},\zeta) =&& \frac{1}{c(\zeta)}e^{-\frac{\kappa\xi^2}{c(\zeta)}}\bigg[\sum_{l=1}^5\alpha_le^{-i\frac{\chi_l^2\zeta}{2c(\zeta)}}J_0\Big(\frac{\chi_l\xi}{c(\zeta)}\Big)\nonumber\\
&&+\sum_{l=1}^3\beta_le^{-i\frac{{\chi'}_l^{2}\zeta}{2c(\zeta)}}e^{2i\phi}J_2\Big(\frac{\chi'_l\xi}{c(\zeta)}\Big)\nonumber\\
&&+\sum_{l=1}^3\gamma_le^{-i\frac{{\chi''}_l^2\zeta}{2c(\zeta)}}e^{-2i\phi}J_{-2}\Big(\frac{\chi''_l\xi}{c(\zeta)}\Big)\nonumber\\
&&+\delta_1e^{-i\frac{\chi^2\zeta}{2c(\zeta)}}e^{4i\phi}J_4\Big(\frac{\chi\xi}{c(\zeta)}\Big)\nonumber\\
&&+\delta_2e^{-i\frac{\chi'^2\zeta}{2c(\zeta)}}e^{-4i\phi}J_{-4}\Big(\frac{\chi'\xi}{c(\zeta)}\Big)\bigg].
\label{fhopfp}
\end{eqnarray}

As one can see, the procedure can formally be continued in a systematic way. It should be noted, however, that the complexity of this expression has significantly increased. In order to generate such a knot, 13 BG beams are now needed! Unlike this, for the trefoil or the Boromean rings, 5 beams were sufficient. Thus, producing more and more complicated knots proves to be quite a challenge. It is a problem from the numerical point of view as well, since with many nodal lines passing close to one another, extremely high precision of calculations is required, which, combined with the large number of terms in the formula, results in long computer runtime.

\section{Summary}\label{sum}

The creation of knotted vortex lines in light beams is one of the most interesting challenges in topological optics. The current paper presents a systematic method of obtaining superpositions of Bessel-Gaussian mods such that the nodal lines (where field intensity drops to zero) form given knot structures. We have selected five geometries as examples: the unknot, the Hopf link, the Borromean rings, the trefoil and the figure-eight knot. It is quite a wide spectrum of choice which indicates a certain universality of the approach. However, with the increasing complexity of a knot, a growing number of BG wave components were needed. In the case of the unknot, $3$ were sufficient; for the most complex knot (figure-eight), $13$ were needed. However, because of the fact that BG beams are relatively easy to obtain in experiments, one can imagine entangled traps for particles obtained by means of these waves.

The proposed method is systematic: in principle, it can be relatively easily applied to increasingly complex structures. Yet, it must be understood that these superpositions will comprise a very large number of component waves. This is reflected in the numerical calculations such as those presented in this paper, where computer time dramatically increases for complex superpositions. There is also another reason for this increase: for complex knots several nodal lines are passing close to one another, which requires high accuracy. This was demonstrated by comparing the Borromean rings and the trefoil.

Using a different representation of the Schr\"odinger propagator than that given in formula (\ref{propagator}), the proposed method can readily be applied to other types of beams. Moreover, it seems that the nodal lines obtained from this method are much smoother and contain fewer ``fringes'' (naturally with identical topology) than those formerly produced with other methods. This may be of practical importance, as keeping the particles in such traps or guiding them along knotted lines presents some kind of a real challenge, and the smoother the trapping-potential valley is, the easier is to achieve it.

\end{document}